\documentclass[amsmath,superscriptaddress,amssymb,aps]{revtex4}

\usepackage{graphicx}
\usepackage{soul}
\usepackage[colorlinks=true,citecolor=blue,linkcolor=magenta]{hyperref}
\usepackage[usenames]{color}
\usepackage{setspace}

\usepackage[UKenglish]{babel}
\usepackage{amsmath} 
\usepackage{amssymb}

\newcommand{\ket}[1]{\left\vert #1 \right\rangle}

\def \ket#1{|#1\rangle}

\def \be{\begin{equation}}
\def \ee{\end{equation}}
\def \ba{\begin{array}}
\def \ea{\end{array}}
\def \bea{\begin{eqnarray}}
\def \eea{\end{eqnarray}}

\renewcommand{\phi}{\varphi}

\usepackage{tabularx,ragged2e,booktabs}
\newcolumntype{C}[1]{>{\Centering}m{#1}}

\newcommand{\hc}{\mathit{h.c.}}

\begin{document}

\title{Resolving single molecule structures with Nitrogen-vacancy centers in diamond}

\author{Matthias Kost ({\it matthias.kost$@$uni-ulm.de})}
\affiliation{Institut f\"{u}r Theoretische Physik, Albert-Einstein Allee 11, Universit\"{a}t Ulm, 89069 Ulm, Germany}
\affiliation{Center for Integrated Quantum Science and Technology, Universit\"{a}t Ulm, 89069 Ulm,
Germany}
\author{Jianming Cai}
\affiliation{Institut f\"{u}r Theoretische Physik, Albert-Einstein Allee 11, Universit\"{a}t Ulm, 89069 Ulm, Germany}
\affiliation{Center for Integrated Quantum Science and Technology, Universit\"{a}t Ulm, 89069 Ulm,
Germany}
\author{Martin B. Plenio}
\affiliation{Institut f\"{u}r Theoretische Physik, Albert-Einstein Allee 11, Universit\"{a}t Ulm, 89069 Ulm, Germany}
\affiliation{Center for Integrated Quantum Science and Technology, Universit\"{a}t Ulm, 89069 Ulm,
Germany}
\date{\today}

\begin{abstract}
We present theoretical proposals for two-dimensional nuclear magnetic resonance spectroscopy
protocols based on Nitrogen-vacancy (NV) centers in diamond that are strongly coupled to the
target nuclei. Continuous microwave and radio-frequency driving fields together with magnetic
field gradients achieve Hartmann-Hahn resonances between NV spin sensor and selected
nuclei for control of nuclear spins and subsequent measurement of their polarization dynamics.
The strong coupling between the NV sensor and the nuclei facilitates coherence control of nuclear
spins and relaxes the requirement of nuclear spin polarization to achieve strong signals and
therefore reduced measurement times. Additionally, we employ a singular value thresholding
matrix completion algorithm to further reduce the amount of data required to permit the
identification of key features in the spectra of strongly sub-sampled data. We illustrate
the potential of this combined approach by applying the protocol to a shallowly implanted
NV center addressing a small amino acid, alanine, to target specific hydrogen nuclei
and to identify the corresponding peaks in their spectra.
\end{abstract}

\maketitle

\section*{Introduction} Nuclear magnetic resonance spectroscopy (NMR) allows for the structure
determination of molecules and proteins and therefore contributes fundamentally to the advancement
of the biological sciences. This structural information is obtained by probing the target of
investigation, typically a large molecular ensemble, by means of multiple radio frequency pulses
and measuring their response. This response is then mapped to multi-dimensional spectra which
encode the dynamical properties of the system and therefore of the interactions between its
constituent nuclear spins \cite{Jeener, AueBE76, Ernst89}. These data in turn permit the reconstruction
of their mutual distances and from this of the entire molecular structure. Due to the minute size
of the nuclear magnetic moments compounded by the tiny polarization of these nuclear spins at
room temperature, even in very strong magnetic fields, large ensembles (at least $10^{12}$ molecules)
are required in order to extract a measurable signal. Thus NMR is susceptible to inhomogenous
broadening and can only deliver ensemble information while the structure and dynamics of individual
specimens remain hidden from observation \cite{Barkai04, Deniz08,Lord10}.

The recent progress in the control of a single electron spin in Nitrogen-vacancy (NV) centers in
diamond offers a new perspective here as it becomes possible to use optically detected magnetic
resonance \cite{Koehler93,Wrachtrup93} to read out the effect of smallest magnetic fields \cite{Chernobrod05,Degen08,Maze08,Bala08}.
Recent theoretical investigations \cite{Cai13,Viktor13} have suggested that shallowly implanted
NV centers \cite{Degen12,Stau12} in conjunction with dynamical decoupling methods
\cite{NaydenovDH+11,CaiJK+12,CaiNP+12,WangCR+14} should be able to detect and locate individual
nuclear spins above the diamond surface. Subsequent experimental work has achieved the observation
of ensembles of nuclear spins outside of diamond \cite{Stau13,Mamin13,Walsworth14} and more recently
the detection of single digit numbers of silicon nuclear spins with a sensitivity that is sufficient
to identify even individual nuclear spins \cite{Mueller14}. Remarkably, this experiment has also
demonstrated a new detection regime in which the NV center (that is a quantum sensor) couples more
strongly to the external targetted nuclear spins than these spins are coupling to their neighbours.
In consequence, even an unpolarised sample can lead to a signal with full contrast as the random
nuclear spin flip-flops are to slow to average out the NV-nuclear spin interaction. This alone
results in a million-fold improvement of sensitivity over standard NMR \cite{Mueller14}.

Beyond this remarkable enhancement of the NMR signal, it is important to realize that this new
detection regime offers further opportunities beyond the capabilities of standard NMR. In addition
to the manipulation and probing of nuclear spins by means of external radio-frequency fields we are
now in a position to control the properties of the NV center such as to tailor its response and,
crucially, via its strong interaction with nuclei also to obtain an additional handle for the manipulation
of the nuclear spins. In this work we take advantage of this potential and demonstrate the usefulness
of single molecule NMR by means of strongly coupled NV centers.

The large amount of required data and the associated long measurement times represent a challenge
that is common to both ensemble NMR and single molecule NMR measurements. However, the spectral
information that is being obtained in any NMR experiment possesses underlying structure, determined
by the Hamiltonian that describes the mutual interactions of the nuclear spins in the molecule, which
makes the system sparse in a suitable basis. This fact can be exploited by non-linear reconstruction
methods from signal processing that are designed to unravel such structures without prior knowledge
with the minimal possible number of measurements. Indeed, measuring only a small subset of all
accessible data points can be shown to allow for the reliable reconstruction of spectral information
as it is required in NMR protocols.

In this work we are combining one such approach, matrix completion \cite{CandesW08,JFC10} (see
\cite{HollandBG+11} for applications of the related but distinct compressive sensing to bulk NMR),
and NMR spectroscopy with an NV spin quantum sensor in the strong coupling regime to devise a new
regime of single molecule NMR that may provide a novel route towards the long-term goal of
elucidating the structure of individual molecules and proteins.

This article proceeds as follows: The first section begins with a numerical discussion of single
molecule two-dimensional 2D-NMR (COSY) by means of strongly coupled NV centers in diamond. It also
introduces techniques of selective polarization enabled by strong coupling to enhance the contrast
in the NMR spectra and presents realistic numerical simulations of the envisaged set-up. A second
part of this section presents protocols that employ an NV quantum sensor that is strongly coupled
to external nuclei to resolve inter-molecular couplings in 2D correlation spectroscopy at the
single molecule level even without previous polarisation of the target molecule. For illustration
we apply these schemes numerically to simulations of the application of our scheme to a simple
bio-molecule, alanine.
In a final section this article discusses and demonstrates the power of matrix completion to reduce
the data taking effort and hence experiment time in single molecule NMR by means of numerical examples.

\section*{Basic principles of 2D NV spectroscopy in the strong coupling limit} \label{theory_NMR}
In this section, we will introduce the basic principles of 2D correlation spectroscopy (COSY)
which we will then use as a test case for the theoretical study of the exploitation of the
strong coupling regime of an NV quantum sensor with target nuclei.

\begin{figure}[t]
    \centering
    \includegraphics[width=0.6\linewidth]{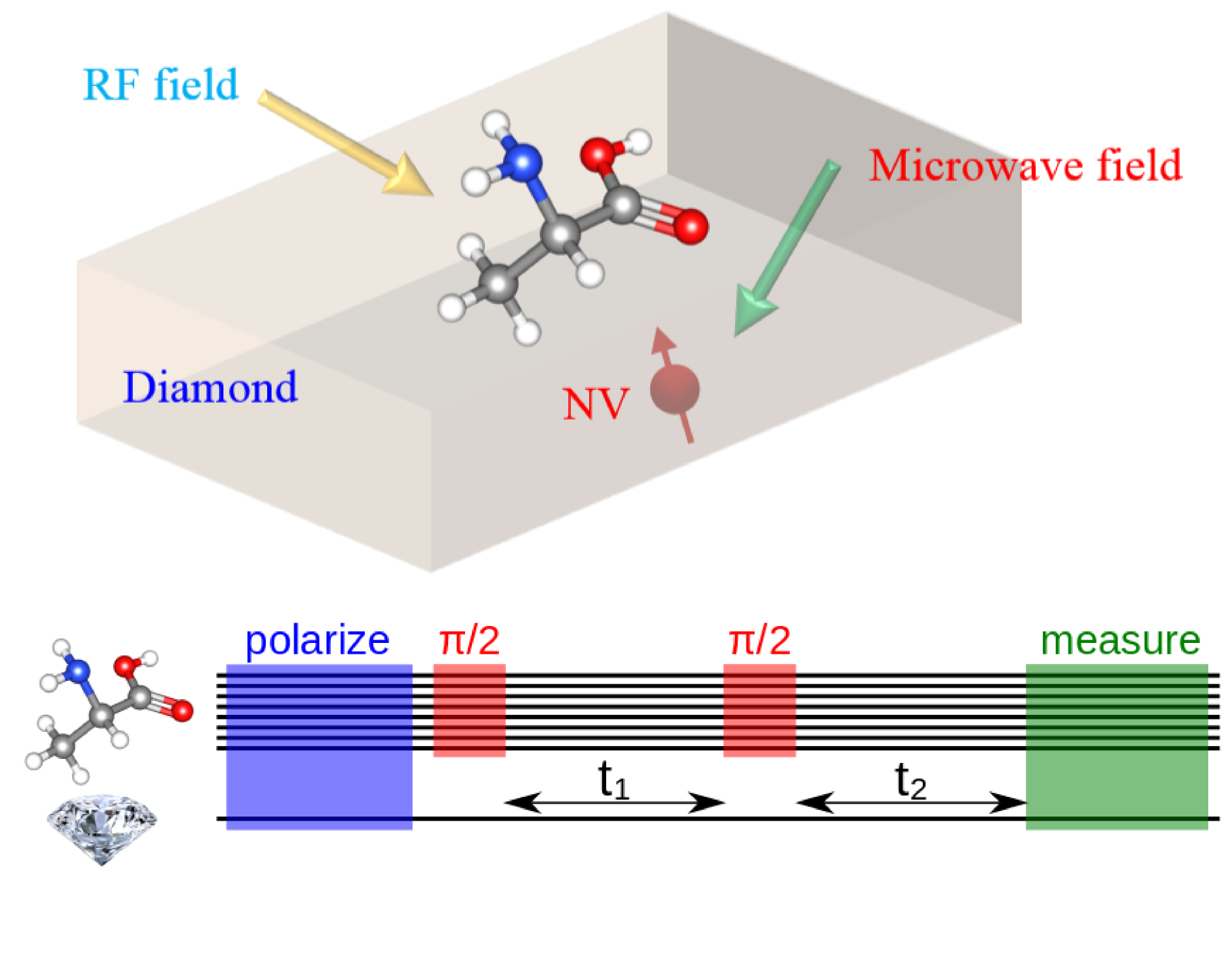}
    \caption{Two-dimensional correlation spectroscopy (COSY) pulse sequence. The nuclei (upper row of horizontal lines)
        are initially polarized by a NV spin sensor (blue box) and prepared into a coherent superposition by the subsequent
        application of a radio-frequency $\pi/2$-pulse (red box) on the nuclei only. The nuclei then undergo a free evolution
        for time $t_1$ followed by a second radio-frequency $\pi/2$-pulse (red box). After a further free evolution time
        $t_2$, the polarization of nuclear spins is measured by a NV spin sensor (green box). During the free evolution
        times, the interaction between the NV center electron spin and the nuclei is eliminated by transferring the NV
        spin to the $m_s=0$ ground state.}\label{img_pulsescheme}
\end{figure}

{\em Elements of COSY in the strong coupling limit --} The basic variant of COSY employs two pulses
separated by an incremented delay $t_1$ where the response of the system is measured after a time
interval $t_2$ (see red and green sections of Figure \ref{img_pulsescheme}). The experimental data
collected in this manner is typically Fourier transformed in both dimensions ($t_1$ and $t_2$) to
yield the frequency domain COSY spectrum. The resulting diagonal peaks of the signal in frequency
space refer to the eigenfrequencies of the nuclear spin system, and the off-diagonal peaks indicate
polarization exchange between pairs of nuclei in the target molecule.

In COSY, as is common to standard NMR protocols executed on large ensembles or in the weak coupling
regime, the measured signal will be proportional to the nuclear spin polarisation. In the strong
coupling regime we can enhance the collected signal by making use of the possibility of achieving
hyperpolarization, i.e. nuclear spin polarization far beyond thermal equilibrium conditions, of all
or a selected few of the nuclei by means of dynamical nuclear spin polarization via the NV centers
\cite{London13,Qiong15}. Hyperpolarization of specific sets of nuclei can be achieved by selection
in frequency space via magnetic field gradients, together with radio-frequency pulses (summarized
as the blue shaded part in Figure \ref{img_pulsescheme}). We remark that it is possible to avoid
the use of radio-frequency pulses by leveraging the strong coupling between the NV spin sensor and
the nuclei to achieve indirect control over the nuclei by means of microwave control of the NV
center electron spin \cite{ShiKW+13}. The measurement of nuclear spin polarization will finally
be achieved by the same NV quantum sensor \cite{CaiNatPhy13}.

Neglecting rapidly oscillating terms, that is under the rotating wave approximation, the dynamics
of the NV center and the nuclei in an interaction picture with respect to $H_0 = \omega_{MW}\sigma_z$,
where $\hbar=1$ and $\omega_{MW}$ is the frequency of the microwave drive, is governed by the effective
Hamiltonian
\begin{eqnarray}
    H_p^{(2)} = \frac{\Omega_{nv}}2 \sigma_x + \omega_f\sum\limits_i\mathbf{ s}^z_i
    + \sum\limits_i g_i^{\parallel}\mathbf{ s}^z_i + \sum\limits_ig_i^{\perp}\left( \sigma_{x}^+
    \mathbf{ s}^-_i + \hc \right). \label{eq_NV_nuclei}
\end{eqnarray}
Here $\Omega_{nv}$ is the effective Rabi frequency of the microwave field applied to the electronic
$\ket{m=-1}\leftrightarrow \ket{m=0}$ transition between of the NV spin in the electronic
ground state, and $\sigma_x$ is the Pauli operator in the corresponding subspace spanned
by $\{ \ket{m=-1}$, $\ket{m=0}$\}, $ \sigma_{x}^+$ is the raising operator in the eigenbasis
of $\sigma_x$ and $\mathbf{s}_i^{-}$ represents the nuclear spin lowering operators. Nuclear
spin polarization can now be achieved by exploiting a Hartmann-Hahn condition between the
Rabi frequency of the driven electron spin of the NV center and the Larmor frequency of the
nuclear spins \cite{London13}. It has been observed before \cite{CaiNatPhy13} that the polarization
procedure can be enhanced if the internuclear coupling is reduced during the polarization and
the measurement process. This can be achieved by applying radio frequency fields which are most
effective for a detuning $\Delta_p$ from the Larmor frequency of the nuclei and a field strength
$\Omega_p$ that satisfy $\Omega_p = \sqrt2\Delta_p$. The result is a new effective nuclei energy
scale $\omega_f = \sqrt{\Delta^2_p + \Omega^2_p}$ \cite{CaiNatPhy13}. In this case the Hartmann-Hahn
condition changes as it needs to account for the detuning and driving fields to find
\begin{equation}
    \Omega_{nv} = \left(\gamma_N B - \Delta_p\right)+ \omega_f.
    \label{Condition}
\end{equation}
If this conditions is fulfilled, electron-nuclear spin-flip-flop processes lead to polarization
exchange between the NV center spin and the nuclei and result in the efficient polarization of
the nuclear spins. Alternatively to the radio-frequency field decoupling, a strong magnetic gradient
may be applied in which case even identical nuclear spin species will have different Larmor frequencies.
While this may require a sweep across the frequency range, as implemented e.g. in the integrated
solid effect \cite{Henstra88,Qiong15}, it also opens the opportunity to allow polarisation to
dominate for specific nuclei.

Finally, the measurement of the polarization of the nuclei is also achieved by the NV center. Indeed,
we can infer the average polarization of the nuclei from the measurement signal of the NV center
spin by \cite{CaiNatPhy13}
\begin{equation}
    \Delta_P = P^+_- - P^-_+ = 2\tau^2 \sum\limits_i(g_i^{\perp})^2\langle\mathbf{s}^z_i\rangle
\end{equation}
where $P_\mu^\nu (\mu,\nu = \pm$) is the population of the NV center in state $\ket\nu$ after being
initialized in state $\ket\mu$ and a free evolution determined by eq. \eqref{eq_NV_nuclei} for a period
of time $\tau$. The nuclear spin polarization in $x$ and $y$ direction can be measured similarly by
applying an additional pulse to rotate the nuclear spins before performing the NV measurements
\cite{CaiNatPhy13}.

We demonstrate the basic working principle of this protocol at the hand of the simple example
of two Hydrogen atoms at a distance of $1 {\mathring{A}}$, which represents a lower limit to
the distance between hydrogen in proteins. In our numerical simulation, we assume an applied 
magnetic field of $1000$G which
allows us to comfortably resolve dipole-dipole energy splitting of about $100$ kHz for our
choice of physical simulation times for a single measurement of a few milliseconds. The NV
center is assumed to be located at a distance of $2$ nm to both Hydrogen nuclei. To support selective addressing of the nuclei, the
applied magnetic field includes a gradient of $60$G/nm along the axis joining the Hydrogen atoms.
This allows for a dominant polarization of a distinct nucleus in the sample, dominant coupling
of the NV center to a specific nucleus during the read out and a maximum fidelity of applied
pulses for selected nuclei. The applied $\pi/2$-pulses are implemented by radio-frequency fields
with a Rabi frequency of $5$ kHz.

\begin{figure}[t]
    \centering
    \includegraphics[width=0.95\linewidth]{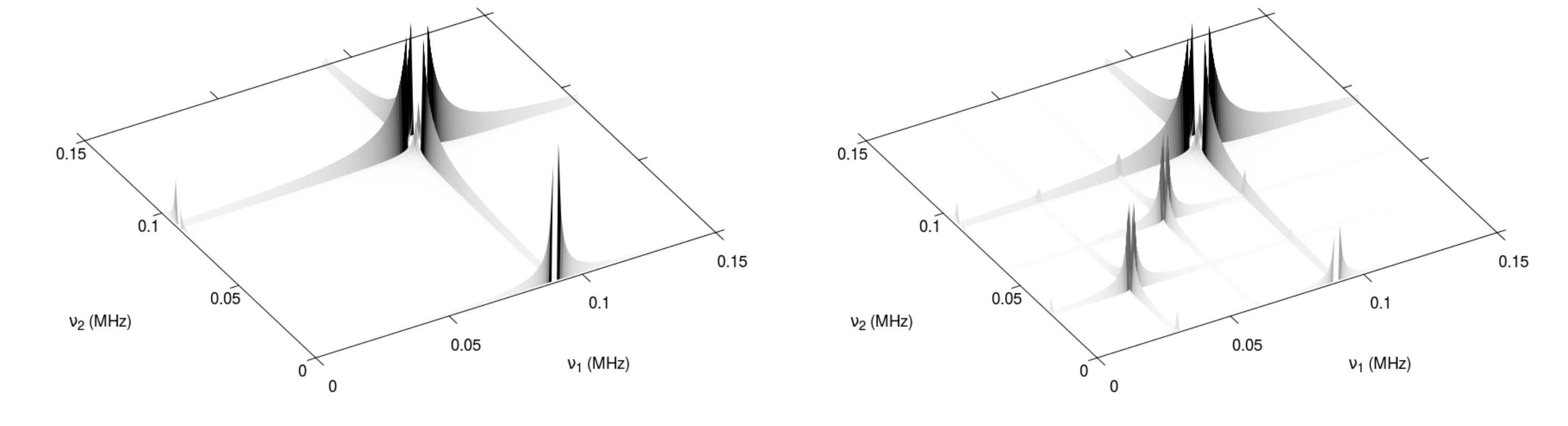}
    \caption{Two-dimensional correlation spectroscopy (COSY) simulation of two Hydrogen atoms arranged
    at a distance of $1 {\mathring{A}}$. As described in the main text, this highly symmetric sample is
    disturbed by an additional Nitrogen atom, which couples only to one of the two nuclei. {\bf Left:}
    no application of selection mechanisms during the protocol. Polarization signal is averaged out
    over the sample. {\bf Right:} The application of selective polarization and control reveals NMR
    information from single nuclei beyond sample averaged quantities. Features such as the new peaks
    around $0.03$ MHz and $0.06$ MHz have also been verified by further simulations that measure the
    nuclear polarization by direct projection rather than invoking an NV center as sensor.
    }\label{img_compare_small_example}
\end{figure}

Figure \ref{img_compare_small_example} compares the result of applying the above COSY protocol
without (l.h.s.) and with (r.h.s.) the selection mechanisms. The latter case exhibits additional
peaks which emerge due to the selective polarization of one of the Hydrogen nuclei. It should be
noted that the protocol includes the internuclear coupling (of order $100$ kHz) which exceeds
the energy shift due to the applied magnetic field gradient (of order $25$ kHz) but is nevertheless
sufficient to result in clearly identifiable additional peaks in the spectrum. This confirms nicely
our expectations, because dropping the selection mechanisms, one may only gain average information
for the whole sample. Further simulations suggest, that selective coupling can also lead to effects,
where spectra become less crowded by suppressing the unwanted influence of certain nuclei during the
read out as will be discussed in more detail for a larger molecule, alanine.
In the setting presented so far, we do not identify yet the orientation of the Hydrogen molecule
relative to the NV-coordinate system. This can be achieved by exploiting the dependence of the
energy splitting due to the internuclear interaction on the relative orientation w.r.t. the
externally applied magnetic field as given by
\begin{equation}
   \delta_{dd} = \frac{\mu_0}{4\pi}\frac{\hbar\gamma_a\gamma_b}{r^3} (3\cos^2\theta-1). \label{eq_angledep}
\end{equation}
Here $r$ is the distance between two nuclei, $\cos\theta = \hat b\cdot\hat r$ with $\hat b$
and $\hat r$ the unit direction vector of the magnetic field and the vector that connects two
nuclei respectively and $\gamma_a,\gamma_b$ are the nuclear gamma factors.
Varying the magnetic field orientation (with three possible directions), one can extract the
relative orientation of pairs of nuclei in the sample by fitting Eq. \eqref{eq_angledep} to
the splitting of matching pairs. To obtain the values of $r$ and $\theta$, two independent magnetic
field directions are sufficient. From the distances and orientations of the various nuclei relative
to the selectively addressed nucleus in the sample, we can learn the spatial structure of its
local environment within the molecule.

{\em Single molecule NMR of Alanine --}
Before moving on, we consider the NV-based COSY spectrum of an Alanine molecule in the strong
coupling regime as described in the above section. Alanine is one of the smallest amino acids
with chemical composition HOOCCH(NH${}_2$)CH${}_3$, which carries nuclear spins on the $7$
Hydrogen atoms and the single Nitrogen atom. The molecule has a total size of about $0.45$nm
and is illustrated in the upper part of Figure \ref{img_pulsescheme}. The nuclear coordinates
and the internuclear coupling rates can be found in the Supplemental Information.

We assume a shallowly implanted NV center located at a distance of $2$nm to the surface \cite{Mueller14}
to measure the polarization of the Hydrogen and Nitrogen atoms by matching the resonance condition
for the two species respectively. In this manner we can obtain the corresponding magnetic resonance
spectra. Apart from the fact that in the strong coupling regime the NMR signal is enhanced for single
molecules the combination with magnetic field gradients opens the possibility to isolate specific
parts of the NMR spectrum. Indeed, to further distinguish between specific Hydrogen atoms, we suggest
to apply a combination of a gradient field and an external radio-frequency drive to support selective
polarization and coupling to individual single Hydrogen atoms by suitable tuning of the NV center
to the corresponding transition frequency. This will be particularly effective if the energy mismatch
between the Larmor frequency of the NV center and the frequency of the not-to-be-addressed
Hydrogen atoms exceeds their coupling strength. For neighboring Hydrogen nuclei in Alanine their 
distance is of order $1.8 {\mathring{A}}$ which results in an interaction of order of around $20$ 
kHz. While it is difficult to exceed this by means of magnetic field gradients our simulations 
show that the combination of a rf-field with Rabi frequency of $100$ kHz with a realistic gradients 
of $60$G/nm \cite{Mamin12} (resulting in energy shifts in the range $[0,45\mbox{kHz}]$ depending on 
the relative orientation of the magnetic field gradient to the positions of the relevant nuclei) 
we are able to observe significant selective polarisation and coupling.

\begin{figure}[t]
    \centering
      \includegraphics[width=0.9\linewidth]{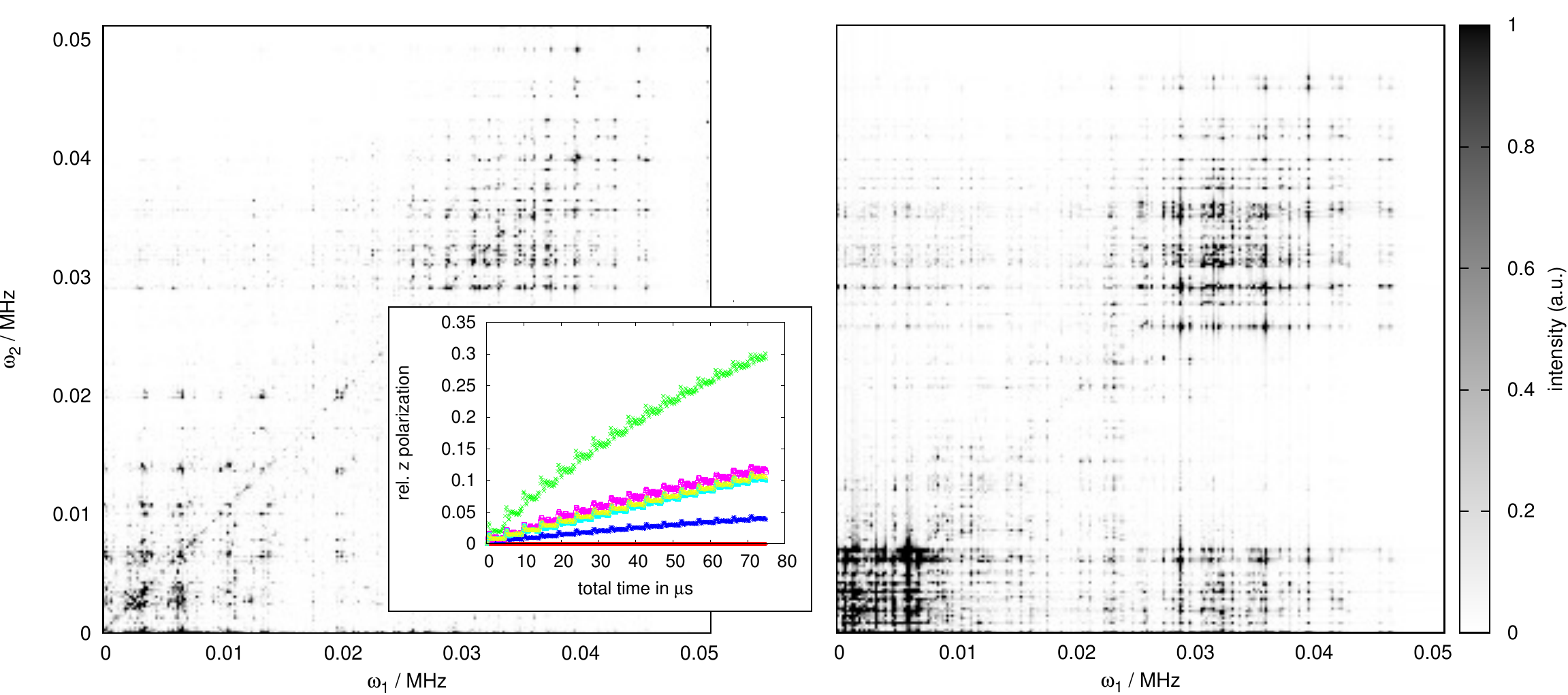}
      \caption{ 2D COSY NMR spectrum of alanine. {\bf left:} including selective polarization
      as described in the main text. The splitting occurs due to the hyperfine interaction of
      the selected nuclear spin with its neighborhood. The peak coordinates are given with respect
      to the Hydrogen Larmor frequency. {\bf inset:} individual polarization gain for six of
      $8$ nuclear spins in alanine during the initial preparation sequence. The frequency and
      gradient parameters have been optimized to favour addressing of a specific Hydrogen nuclear
      spin. Due to the strong interaction between the nuclear spins, nuclear polarization transfer
      can not be completely suppressed. However, a two to three-fold increase in polarization of
      the selected nuclear spin as compared to the non-selected ones will result in a remarkable gain
      in contrast. The same selective coupling efficiency can be expected to hold also during other
      parts of the protocols that are based on the selective coupling mechanism.
      {\bf right :} The signal resulting from a fully polarized molecule due to a lack of selective
      addressing. The spectrum is considerably more congested, clearly highlighting the advantage
      of dominant addressing of selected nuclear spins.}
      \label{img_alanine_res}
\end{figure}

In order to achieve dominant coupling to a selected hydrogen atom we perform our selective polarization
protocol and apply the COSY procedure with the NV center tuned on resonance to the hydrogen transition
frequency. This yields the l.h.s of figure \ref{img_alanine_res}, where the hyperfine coupling can be
clearly identified. The decongestion of the spectrum thanks to the dominant addressing of selected nuclei
can be observed very clearly by a comparison to a spectrum that one obtains without selective addressing
from a fully polarized molecule (see r.h.s of figure \ref{img_alanine_res}). The ability to selectively
address a single Hydrogen spin enables the implementation of the NMR protocol for selected atoms in the
target molecule and therefore allows for the reconstruction of the molecular environment of the chosen
atoms by reading out the hyperfine splitting of each atom individually.

The results of this simulation will be used also for the demonstration of the application of matrix
completion in the last section.

\label{theory_StrongCoupling}
{\em Exploiting entanglement between the NV electron spin and the target nuclei --} The protocols
so far exploit the ability to achieve hyperpolarization of the molecule or of some of its constituents.
Importantly, however, operating in the strong coupling regime, it is not strictly necessary to polarize
nuclear spins, while still retaining a full signal contrast due to the quantum nature of the interaction
between the NV center spin and the nuclei. Moreover, we can make use of the strong coupling for
the coherent control of both nuclear spins (by means of radio-frequency fields or via the interaction
with the NV center) and of the NV center spin (microwave frequency fields) thus allowing for more
complex pulse sequences for the simultaneous control of sensor and target.

As an illustrative example, we construct a novel NV-based 2D-spectroscopy scheme by exploiting the
strong interaction of nuclei in a target molecule with an NV center spin, see Figure \ref{Strong_coupling_seq}(a).
Here we selectively trigger the NV-nuclear spin interaction to generate entanglement between the NV
center and the nuclear spins interspersed with free evolution times $t_1$ and $t_2$ which in turn
allow for an accumulation of the effect of the nuclear Hamiltonian that can subsequently be measured
via the NV center. While this scheme appears similar to the well-known COSY sequence it differs
from it due to the entanglement that is created between NV center and nuclear spin which in turn
allows us to generate the same spectral information as in the case of COSY with fully polarized
nuclear spins but now without the need for nuclear spin polarization. Although the actual detection
protocol is quite different, this scheme shares sufficient parallels to the COSY scheme from figure
\ref{img_pulsescheme} to transfer the interpretation of peak positions from the original COSY scheme.

For definiteness, we describe the pulse sequence in detail together with the relevant Hamiltonians
 (see also fig. \ref{Strong_coupling_seq}):

\begin{figure}[t]
    \centering
    \includegraphics[width=0.6\linewidth]{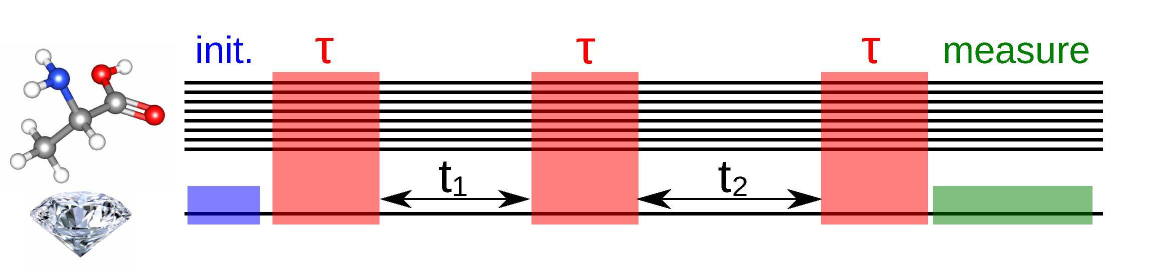}
\caption{ 2D NMR spectroscopy pulse sequence with an NV sensor in the strong coupling regime.
The nuclei are initially unpolarized, only the NV center is initialized in a $m=\pm1$ state.
The final measurement is a projection of the NV center onto its initial state. During the
free evolution times $t_{1,2}$, the NV-nuclear interaction is effectively eliminated, while
it is switched on during the periods $\tau$. }
\label{Strong_coupling_seq}
\end{figure}

\begin{itemize}
 \item {\it Initialization}: the NV center is initialized in a polarized state, while the nuclear spins remain unpolarized (blue bar).
 \item {\it NV-nuclear interaction}: the NV center interacts with the nuclei for a time $\tau$ (red bar). In this step the NV center becomes entangled with the nuclear spins. The dynamics is governed by the Hamiltonian eq. (\ref{eq_NV_nuclei}), where the additional application of an external RF fields effectively decouples the nuclear spins as explained above and again leads to the resonance condition given in equation \eqref{Condition}.
 \item {\it Free evolution period 1}: the interaction of the NV center with the nuclei is switched off by transfer of the NV center to the $m=0$ state
    and if necessary the quantum information may be transferred further to the nuclear spin degree of freedom of the NV center. The nuclear spins precess
    freely for a time $t_1$ during which the dynamics is governed by the Hamiltonian
     \begin{eqnarray}
        H_B = \sum\limits_i\gamma_i {\bf B}\cdot {\bf s}_i
        + \sum\limits_ {i\ne j} g_{ij} \left( {\bf s}_i \cdot {\bf s}_j - 3(\hat {\bf r}_{ij} \cdot {\bf s}_i)(\hat {\bf r}_{ij} \cdot {\bf s}_j) \right)
     \end{eqnarray}
\item {\it NV-nuclei interaction}: The NV center again interacts with the nuclei with the same Hamiltonian eq. (\ref{eq_NV_nuclei}).
    The additional entanglement depends on the system state after the previous free evolution time $t_1$.
\item {\it Free evolution period 2}: the interaction with the NV center is switched off and the nuclear spins precesses freely for a time $t_2$ under the same Hamiltonian as in the free evolution period 1.
\item {\it NV-nuclear interaction}: A third interaction employing the same dynamics as in the previous two NV-nuclear
interaction periods prepares the final measurement.
\item {\it Measurement}: Finally, we perform a projective measurement on the initial state of the NV center spin (green bar).
\end{itemize}
\begin{figure}[t]
\includegraphics[angle=-90,width=0.6\linewidth]{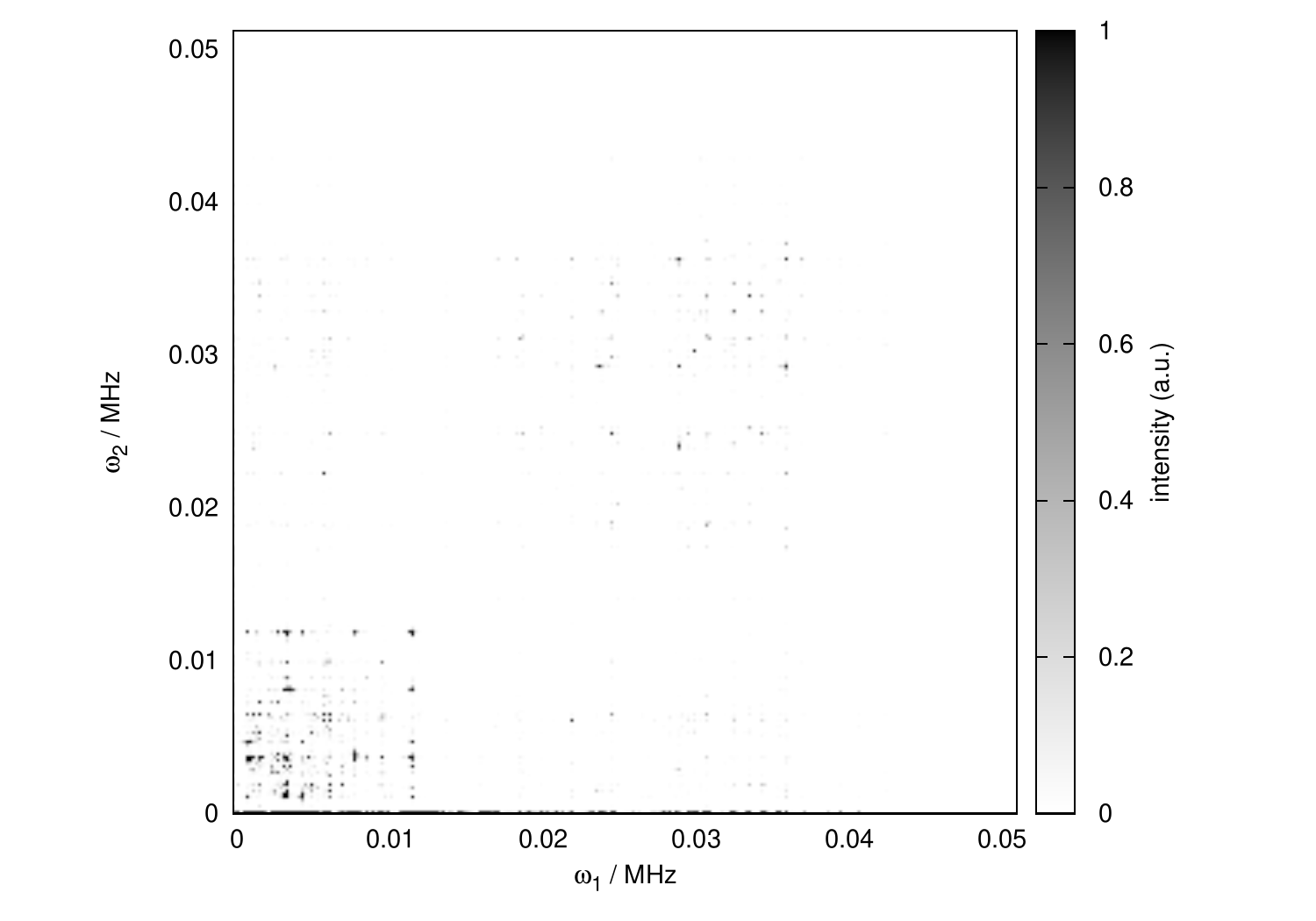}
\caption{ Entanglement based 2D NMR spectroscopy pulse sequence with an NV sensor in the strong coupling regime.
Resulting spectrum for an NV center tuned on resonance to a selected Hydrogen transition at a magnetic field of
100G aligned to the NV axis. $t_{1,2}$ was varied up to 5ms, while the time $\tau$ has been chosen around 1.25ms.
Due to the long simulation times, the hyperfine coupling is resolved. The selection mechanism was tuned to another
nucleus in this simulation than the COSY equivalent in figure \ref{img_alanine_res}, which demonstrates the
distinguishability.}
\label{Strong_coupling}
\end{figure}

As illustration of the above 2D NMR protocol with a strongly coupled NV center at a depth of 2 nm we
consider Alanine in a magnetic field of $100$G. To preferably address selected nuclear spins we apply
an additional gradient field of $60$G/nm, to support dominant entanglement between the NV and selected
nuclear spins during the NV interaction periods. The simulated magnetic resonance spectrum is plotted
in Figure \ref{Strong_coupling} from which one can clearly identify the hyperfine splittings. The spectrum
is decongested thanks to the dominant coupling to a specific Hydrogen atom supported by strong external
magnetic field gradients. This selective coupling is particularly useful for measuring hyperfine splittings
of individual atoms, as the coupling strength of the NV center to each Hydrogen atom is nearly identical
in the absence of a gradient field and makes more difficult the distinction of the various peaks in the
spectra. Therefore, selective coupling to each single atom in the molecule allows for the reconstruction
of the molecular neighborhood of atoms and supports the determination of the geometric structure of the
molecule. Without individual atomic addressing, the geometric information has to be extracted from just
a single spectrum, while the ability of selectively addressing $n$ atoms allows for the generation of
$n$ independent spectra to decongest the spectra.

Assuming that the nuclei are initially not polarized, we have also verified that they will remain
nearly unpolarized across the entire pulse sequence, that is we observe NMR signals without relying
on any nuclear polarization, which represents a distinct feature as compared with conventional NMR
techniques. As above, the application of gradient fields allows for individual coupling even for
identical atoms.

\section*{Matrix Completion for an NV-based 2D NMR} \label{matrixcompletion} For larger
molecules, the application of NV-based 2D-NMR suffers from a rapidly increasing experiment
effort as a function of the number of nuclei in the target molecule. In this section, we
explain that the experimental overhead of NV-based 2D spectroscopy can be reduced significantly
by exploiting the technique of matrix completion \cite{CandesW08,JFC10} (see for example
\cite{HollandBG+11} for the related but distinct compressive sensing) which exploits two
specific aspects of NV-based 2D spectroscopy.
First, 2D-spectra generally possess structure which expresses itself in sparseness in a certain
basis and, secondly, while the relevant information is represented in Fourier space, the experimental
data are taken in time. We will begin by clarifying why these two aspects are important
and how they are going to be being used.

{\em Background --} For a given matrix $A$ we denote with $\sigma_i$ its descendingly ordered singular
values, that is the diagonal entries of the matrix D in $A=U\Sigma V^\dagger$ where
$U$ and $V$ are unitaries. If these singular values are close to zero for indices
$i>r$, one obtains a high fidelity approximation $\tilde A$ with rank $r$ for the matrix
$A$ by
\begin{equation}
  \tilde A = U\tilde\Sigma V^\dagger
\end{equation}
where $\tilde\Sigma$ is a low rank version of $\Sigma$, as defined by
\begin{equation}
    \tilde\Sigma_{ij} = \begin{cases} \sigma_i, & i=j\le r \\ 0, & \mathrm{else} \end{cases}
\end{equation}
The idea of matrix completion is to reconstruct the matrix $A$ by finding a low rank
approximation $\tilde A$ based on the knowledge of a few entries from a random sample
set $\left\{(i,j)\right\}=\Omega$ such that $\left\|\mathcal P_\Omega\times(A-\tilde A)\right\|<\epsilon$,
where $\mathcal P_\Omega$ is a boolean matrix that indicates whether $(i,j)$ is in $\Omega$
or not and $\times$ means elementwise multiplication.

It is important to note that generally the basis in which one takes the random samples will
affect the reconstruction efficiency. Indeed, sampling a matrix with just one nonzero entry,
i.e.\ a very sparse matrix, will tend to yield only zeros upon random sampling of the matrix entries
and hence any reconstruction algorithm must conclude that all the entries of the matrix are
in fact zero. This situation differs considerably when we take the discrete Fourier transform
of this matrix and sample the result. Now sampling even a small number of entries will yield
useful information about the structure of the matrix and indeed it becomes possible then to
reconstruct this Fourier transformed matrix (and hence also the original) from a small number
of sampled entries (for a $n\times n$ matrix the number of required samples scales as $rn\ln n$
where $r$ is the singular value rank of the data matrix). Indeed, for a matrix that is sparse
in some basis $\{|e_i\rangle \}$ (i.e.\ most of its entries vanish or are negligible) it is
in this sense optimal to sample the matrix in the Fourier transformed basis $\{{\cal F}|e_i\rangle \}$
(see e.g. \cite{Gross} for a rigorous treatment that provide the mathematical foundation
for these observations).

It is essential that the data matrices that one obtains from NV-based 2D-NMR tend to be
approximately sparse
as their entries are concentrated on the diagonal and around the off-diagonal positions that
indicate coupling between those diagonal elements. From these arguments and observations it becomes
transparent that matrix completion is ideally suited to support NV-based NMR spectroscopy as the
desired information is represented in frequency while the data are taken in time, that is we sample
the desired information in a basis that is Fourier transformed with respect to the information basis.
We will make use of this fact in the remainder of this section.

\begin{figure*}
    \begin{minipage}{0.88\linewidth}
        \centering
        \includegraphics[width=\linewidth]{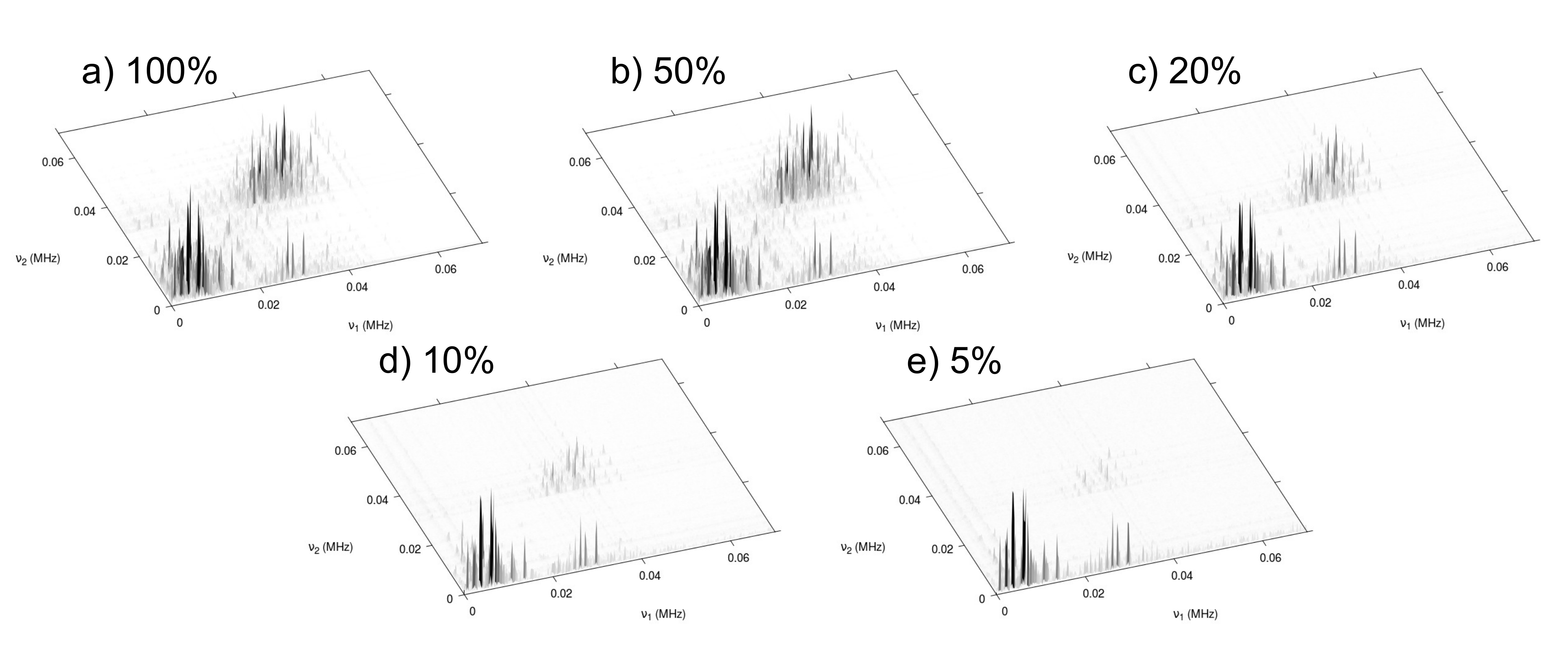}
    \end{minipage}
    \caption{The magnetic resonance spectra of a simulated Alanine molecule based on a NV-based COSY NMR
    and matrix completion.
    Comparison of the distribution of the peaks from the spectrum shown in figure \ref{img_alanine_res} after matrix
    completion for different sample rates. $a$) full data. $b$) $50$\% data sample. $c$) $20$\% data
    sample. $d$) $10$\% data sample and $e$) $5$\% data sample. One can clearly see, how weak contributions
    diminish, while strong contributions remain nearly untouched even at very low sampling rates}\label{img_alanine}
\end{figure*}

{\em Application --}
We simulate a subset of the entries $S_t$ corresponding to randomly chosen $t=(t_1,t_2)$ in the
time-domain signal, which are those data that would be measured in real experiments. The matrix
completion algorithm, whose run time is negligible compared to the savings in measurement time
\cite{Alm12}, will reconstruct the $1024\times 1024$ time domain matrix $S_t$ out of the subsampled
entries. We will see that in the frequency domain, the hyperfine structure can then be identified
from the reconstructed matrix.

As a comparison, we apply the completion algorithm to the Alanine COSY spectra generated above
and plot in Figure \ref{img_alanine}(a) the spectrum obtained from the complete set of measurements,
namely based on a 100\% data sample. In contrast, figure \ref{img_alanine}(b) shows the same as
Figure \ref{img_alanine}(a), for different sample rates. It can be seen that we are able to identify
the relevant splitting structure in the Hydrogen signal even with about $20$\% data sample. Hence
we expect a remarkable time saving due to sub-sampling of both experimental and simulated data for
sufficiently large matrices.

\section*{Conclusions and discussions}
The shallow implantation of NV centers in diamond allow for a new magnetometry regime in which the
coupling of the NV center to nuclei is comparable or larger than the inter-nuclear coupling, in
rapid departure from the standard regime in which nuclear magnetic resonance in bulk is situated.
We have demonstrated that this strong coupling regime offers new opportunities for single molecule
nuclear magnetic resonance spectroscopy that can reduce the experimental requirements both by relaxing
the need for nuclear spin polarization while retaining a significant signal, by decongesting the NMR
spectra thanks to individual addressing for polarization and readout of nuclear spins and by reducing
significantly the number of measurements required to recover the 2D spectrum thanks to the application
of matrix completion, a method from signal processing. In numerical studies we have applied these ideas
to a specific example of a small amino acid, alanine to demonstrate the feasibility and potential of this
approach. We expect that the combination of these approaches may be applied beyond this example to extend
to the study of membrane proteins whose structure and dynamics are of considerable importance in
biology and medicine.

\section*{Acknowledgements} We thank Fedor Jelezko and  Boris Naydenov for discussion
concerning experimental issues and Burkhard Luy for helpful comments on the manuscript.
This work was supported by an Alexander von Humboldt professorship, the DFG SFB TR/21,
the SPP1601, the EU Integrating Projects SIQS and DIADEMS, the EU STREP EQUAM, and the ERC Synergy
grant BioQ.

\section*{Additional Information}

MBP proposed and designed the research, MK carried out the work with support of JMC,
JMC and MBP supervised the project. All authors wrote and reviewed the manuscript
and discussed the research.

{\em Competing financial interests --}
The authors declare no competing financial interests.

\end{document}